\documentclass[a4paper,11pt]{article} 
\usepackage[T2A]{fontenc}			
\usepackage[utf8]{inputenc}			
\usepackage[english]{babel}	
\usepackage{graphicx}
\graphicspath{}
\DeclareGraphicsExtensions{.pdf,.png,.jpg}
\usepackage{amsmath,amsfonts,amssymb,amsthm,mathtools,verbatim,alltt} 
\usepackage{braket}
\usepackage{wasysym}
\usepackage{cancel}
\usepackage{setspace}
\usepackage{slashed}
\usepackage{cite} 
\usepackage[margin=1.5cm]{geometry}
\usepackage{authblk}
\everymath{\displaystyle}
\usepackage{mathrsfs}

\usepackage[colorlinks = true,
linkcolor = blue,
urlcolor  = blue,
citecolor = blue,
anchorcolor = blue]{hyperref}

\date{}
\title{Hydrodynamical duals of the gravitational axial anomaly and the cosmological constant}
\author[1,2,3]{R. V. Khakimov}
\author[1,2]{G. Yu. Prokhorov}
\author[1,2]{O. V. Teryaev}
\author[1,2,4]{V. I. Zakharov}
\affil[1]{Joint Institute for Nuclear Research, Joliot-Curie str. 6, Dubna 141980, Russia}
\affil[2]{NRC Kurchatov Institute, Moscow, Akademika Kurchatova pl. 1, Moscow, 123098, Russia}
\affil[3]{Physics Department, Lomonosov Moscow State University, 1-2 Leninskie Gory, Moscow 119991, Russia}
\affil[4]{Pacific Quantum Center, 
Far Eastern Federal University, 10 Ajax Bay, Russky Island, Vladivostok 690950, Russia}

\begin{document}

\maketitle

\begin{center}
\section*{Abstract}
We construct the hydrodynamic expansion for a rotating and accelerated medium in a curved space-time, and establish a duality between the currents related to the cosmological constant and the acceleration. Then we consider the more general case with a non-zero Weyl tensor, and show the relationship between the current in flat space-time and the gravitational axial anomaly. This generalizes the previous derivation to the case with a non-zero Ricci tensor.

\end{center}

\section{Introduction}

Since the discovery of the chiral anomaly \cite{Adler:1969gk, Alvarez-Gaume:1984} its study has been of great interest. To date, there is a large number of different effects associated with it found in many different physical fields  \cite{Kharzeev:2013jha, Chernodub:2021nff, Fukushima:2008xe, Volovik:2021myq, Bevan1997}. Relativistic fluid effects of the anomalies \cite{Kharzeev:2013jha, Son:2009tf, Chernodub:2021nff, Fukushima:2008xe, Prokhorov:2022udo, Mitra:2021wjp, Gao:2012ix} are of particular interest at the present time, indicating the nontrivial interplay between the infrared and the ultraviolet theories. Besides undoubted theoretical interest, these effects can be applied to describe processes in heavy-ion-collisions \cite{Kharzeev:2022hqz, Rogachevsky:2010ys, Kharzeev:2015znc}.
\\\\
Hydrodynamics, which describes physics at large spatial and temporal distances, is based on the conservation relations, such as conservation of the stress-energy tensor and current \cite{landau2013fluid}
\begin{eqnarray}
 \partial_{\mu} T^{\mu \nu} = 0\,, \quad \partial_{\mu} j^{\mu} = 0\,. 
\label{cons}
\end{eqnarray} 
In particular, in the case of a viscous and heat-conducting fluid the equations (\ref{cons}) allow us to describe linear dissipative effects \cite{landau2013fluid}. A significant step was made in the work \cite{Son:2009tf}, where an additional linear correction associated with vorticity (and magnetic field) was obtained. It was shown that the magnitude of the corresponding vorticity effect (more precisely, the part with chemical potential), so-called chiral vortical effect (CVE), in the axial current
\begin{eqnarray}
j^{\nu}_{A, \text{CVE}} = \left(\sigma_{(T)} T^2 + \sigma_{(\mu)} \mu^2\right) \omega^{\nu} + \mathcal{O}(\partial^3 u) \,,
\label{cve 12}
\end{eqnarray}
is determined by the quantum axial anomaly in the electromagnetic field $F_{\mu\nu}$ \cite{Adler:1969gk}
\begin{eqnarray}
\partial_{\mu}j^{\mu}_A = C\, \varepsilon^{\mu\nu\alpha\beta}F_{\mu\nu}F_{\alpha\beta} \,,
\label{anom gauge gen}
\end{eqnarray}
which, being included, modifies the basic hydrodynamic equations (\ref{cons}).  In (\ref{cve 12}) $ T $ is the proper temperature, $ \mu $ is the chemical potential, $ \omega^{\mu} ={\frac{1}{2}\varepsilon^{\mu\nu\alpha\beta} u_{\nu}\partial_{\alpha}u_{\beta}} $ is the vorticity and $ \mathcal{O}(\partial^3 u) $ denotes terms cubic in gradients of  4-velocity $ u_{\mu} $. The relationship is fixed by the equation, connecting conductivity $\sigma_{(\mu)}$ with a numerical factor from the anomaly $C$
\begin{eqnarray}
\sigma_{(\mu)} = -8 C \,,
\label{theorem1}
\end{eqnarray}
which was obtained in \cite{Son:2009tf} from the second law of thermodynamics for entropy current. In particular, for Dirac field current (\ref{cve 12}) has the form \cite{Kharzeev:2013jha, Zakharov:2012vv, Gao:2012ix}
\begin{eqnarray}
j^{\nu}_{A, \text{CVE}} = \left(\frac{T^2}{6} + \frac{\mu^2}{2\pi^2}\right) \omega^{\nu} + \mathcal{O}(\partial^3 u) \,,
\label{cve 12a}
\end{eqnarray}
which exactly satisfies the well known anomaly for the Dirac field $C=-1/(16 \pi^2)$. In the same way the correspondence (\ref{theorem1}) was demonstrated for spin 3/2 \cite{Prokhorov:2021bbv} and spin 1 \cite{Prokhorov:2020npf}.
\\\\
Recently, it was shown that in a similar way the same effects can be obtained without the use of the entropic current \cite{Yang:2022ksq, Buzzegoli:2020ycf},  taking advantage of the so-called global thermodynamic equilibrium \cite{Becattini:2016stj}.
This simplifies the calculations, which made it possible, in particular, to analyze the second-order gradient effects \cite{Yang:2022ksq}. 
\\\\
Note that the anomalous transport associated with the electromagnetic chiral anomaly was mainly analyzed (\ref{anom gauge gen}).
However, there are also a number of works \cite{Landsteiner:2011cp, Stone:2018zel, Frolov:2022rod, Dolgov:1987xv} that consider the transport associated with the gravitational axial anomaly \cite{Alvarez-Gaume:1984}
\begin{eqnarray}
 \nabla_{\mu} j_{A}^{\mu} = \mathcal{N} \epsilon^{\alpha\beta\mu\nu} R_{\mu\nu\lambda\rho} R^{\quad\lambda\rho}_{\alpha\beta} \,,
 \label{gravanomaly}
\end{eqnarray}
where $\epsilon^{\alpha\beta\mu\nu} = \frac{1}{\sqrt{-g}} \varepsilon^{\alpha\beta\mu\nu} $ is Levi-Civita tensor in a curved space-time and $ \mathcal{N}$ is a numerical constant. 
\\\\
However, there was a problem of direct generalization of the method of \cite{Son:2009tf, Yang:2022ksq} to the gravitational chiral anomaly. Unlike the case with an electromagnetic anomaly (associated with a linear order in gradients), it is necessary to consider higher orders of the gradient expansion. This, as well as the relative complexity of gravity itself, makes a generalization of \cite{Son:2009tf, Yang:2022ksq} somewhat nontrivial and massive. 
\\\\
The solution was recently found in  \cite{Prokhorov:2022udo}, where it was shown that the imprints of the gravitational chiral anomaly (\ref{gravanomaly}) arise in the third order of the gradient expansion for the axial current. Interestingly, this effect exists even in flat space-time, when the corresponding current includes only acceleration and vorticity
\begin{eqnarray}
 j_{\mu}^{A(3)} = \left( \lambda_{1} \omega^{2} + \lambda_{2} a^{2} \right) \omega_{\mu}\,,
 \label{diraccurrent}
\end{eqnarray}
where $a_{\mu}= u^{\nu} \partial_{\nu} u_{\mu} $ is acceleration and $\lambda_2$ and $\lambda_1$ are dimensionless conductivities.
The connection of the current (\ref{diraccurrent}) with the gravitational chiral anomaly (\ref{gravanomaly}) is fixed by the equation
\begin{eqnarray}
 \frac{\lambda_{1} - \lambda_{2} }{32} = \mathcal{N}\,.
 \label{key}
\end{eqnarray}
In particular, for Dirac fields conductivities in (\ref{diraccurrent}) were found directly using the density operator and has the form \cite{Prokhorov:2018bql} (see also \cite{Palermo:2021hlf, Ambrus:2019ayb})
\begin{eqnarray}
 j_{\mu}^{A(3)} = \left(- \frac{\omega^{2} }{24 \pi^2} - \frac{a^{2}}{8 \pi^2}   \right) \omega_{\mu}\,.
 \label{diraccurrent12}
\end{eqnarray}
At the same time, the gravitational chiral anomaly for Dirac fields is well known \cite{Alvarez-Gaume:1984}
\begin{eqnarray}
 \nabla_{\mu} j_{A}^{\mu} = \frac{1}{384 \pi^2} \epsilon^{\alpha\beta\mu\nu} R_{\mu\nu\lambda\rho} R^{\quad\lambda\rho}_{\alpha\beta} \,.
 \label{gravanomaly12a}
\end{eqnarray}
It's easy to see that (\ref{key}) is satisfied.
Similarly, this relation was verified in \cite{Prokhorov:2022snx} for the Rarita-Schwinger-Adler model \cite{Adler:2017shl} containing fields with spin 3/2 (one mode) and 1/2 (two modes). The current was obtained from the quantum-statistical correlators and has the form
\begin{eqnarray}
 j_{\mu}^{A(3)} = \left(- \frac{53}{24 \pi^2} \omega^{2} - \frac{5}{8 \pi^2}  a^{2} \right) \omega_{\mu}\,.
 \label{diraccurrent32}
\end{eqnarray}
At the same time, a gravitational chiral anomaly was found in \cite{Prokhorov:2022rna}
\begin{eqnarray}
 \nabla_{\mu} j_{A}^{\mu} = -\frac{19}{384 \pi^2} \epsilon^{\alpha\beta\mu\nu} R_{\mu\nu\lambda\rho} R^{\quad\lambda\rho}_{\alpha\beta} \,,
 \label{gravanomaly32a}
\end{eqnarray}
Comparing (\ref{diraccurrent32}) and (\ref{gravanomaly32a}), we see that the general expression (\ref{key}) is satisfied again. Thus, there is a new anomalous transport effect in a vortical and accelerated medium (\ref{diraccurrent}), called the kinematic vortical effect (KVE). 
\\\\
However, the derivation in \cite{Prokhorov:2022udo} was not the most general, since it used an assumption about the properties of the manifold.
Namely, Ricci-flat space-times were considered
\begin{eqnarray}
    R_{\mu\nu} = 0\,.
\end{eqnarray}
And the question is, will the obtained key relation (\ref{key}) remain valid beyond this approximation? To answer this question, in this work we will consider a more general case when the Ricci tensor is not equal to zero, but is proportional to the cosmological constant
\begin{eqnarray}\label{rl}
    R_{\mu\nu} = \Lambda g_{\mu\nu}\,,
\end{eqnarray}
which, on the one hand, (if we set the Weyl tensor equal to zero) corresponds to the very important case of (anti-)de Sitter spaces, while from a technical point of view will simplify the derivation, since it leads to additional equations such as $\nabla_{\rho} R_{\mu\nu} = 0$.
\\\\
Similar hydrodynamic expansion of the stress-energy tensor for the medium in the curved space with (\ref{rl}) was recently considered in \cite{Khakimov:2023emy} for particular case with a zero Weyl tensor, where the anomaly (\ref{gravanomaly}) is absent, and accelerated (non-rotating) medium.
In this case, an alternative confirmation for the Unruh effect in curved space-time was obtained, with a temperature depending simultaneously on curvature and acceleration \cite{Deser:1997ri}. Below we will show that the gradient expansion of the axial current also indicates the generalized Unruh effect in curved space.
\\\\
As a warm-up, we start by constructing a hydrodynamic expansion for the axial current in a vortical and accelerated medium in the case when the Weyl tensor is equal to zero.
\\\\
The paper organized as follows. We begin in Section \ref{sec 2} by considering the simplest case of curved space-time with a zero Weyl tensor. The relationship between the KVE and the anomaly cannot be fixed in this case, however,  the relation between the cosmological constant and acceleration will be obtained.
In Section \ref{sec 3} we consider the more general case of Einstein manifolds with nonzero Weyl tensor. In this case we obtain the KVE relation, and confirm the duality between the cosmological constant and acceleration obtained in Section \ref{sec 2}.
In Section \ref{sec 4} we discuss some aspects and implications of our work and give a brief qualitative description of the obtained effects. In the Conclusion we summarize the results obtained.

\section{Duality of the cosmological constant and acceleration}
\label{sec 2}
\subsection{Riemann tensor and gradient expansion}
We consider an uncharged non-dissipative fluid of massless fermions, moving with four-velocity $u_{\mu}(x)$ and characterized by proper temperature $T(x)$, placed in an external gravitational field with the metric $g_ {\mu\nu} (x)$. The system is assumed to be in a state of global thermodynamic equilibrium \cite{Becattini:2016stj}, for which the inverse temperature vector $ \beta_{\mu} = \frac{u_{\mu}}{T} $ satisfies the Killing equation
\begin{eqnarray}
 \nabla_{\mu} \beta_{\nu} + \nabla_{\nu} \beta_{\mu} = 0\,,
 \label{killingeq}
\end{eqnarray} 
which means that we work with the hydrodynamic $\beta$-frame. Note that the condition (\ref{killingeq}) is not some artificial condition and is very close to the known criteria of thermal equilibrium; we will consider this in more detail in the Discussion section. In particular, due to (\ref{killingeq}), we obtain for the second-order covariant derivative
\begin{eqnarray}
 \nabla_{\mu} \nabla_{\nu} \beta_{\alpha} = -R^{\rho}_{\:\:\mu\nu\alpha} \beta_{\rho}\,.
 \label{doubleder}
\end{eqnarray} 
The anti-symmetric combination of covariant derivatives forms the thermal vorticity tensor \cite{Buzzegoli:2017cqy}
\begin{eqnarray}
 \varpi_{\mu\nu} = - \frac{1}{2} \left( \nabla_{\mu} \beta_{\nu} - \nabla_{\nu} \beta_{\mu} \right)\,,
\end{eqnarray} 
which has one vector and one pseudo-vector component, corresponding to (``thermal'') acceleration $ \alpha_{\mu} $ and vorticity $ w_{\mu} $
\begin{eqnarray}
 \alpha_{\mu} = \varpi_{\mu\nu} u^{\nu}\,, \quad w_{\mu} = \frac{1}{2} \epsilon_{\mu\nu\alpha\beta} u^{\nu} \varpi^{\alpha\beta}\,, \quad \varpi_{\mu\nu} = \epsilon_{\mu\nu\alpha\beta} w^{\alpha} u^{\beta} + \alpha_{\mu} u_{\nu} - \alpha_{\nu} u_{\mu}\,.
\end{eqnarray}
Kinematic vorticity $ \omega_{\mu} $ is proportional to the ``thermal'' vorticity $ w_{\mu} $, and the kinematic acceleration $ a_{\mu} $ is proportional to the ``thermal'' acceleration in the state of global equilibrium (\ref{killingeq})
\begin{eqnarray}
 w_{\mu} = \frac{\omega_{\mu}}{T}\,, \quad \alpha_{\mu} = \frac{a_{\mu}}{T}\,.
 \label{trans}
\end{eqnarray}
In the general case the Riemann tensor has the form
\begin{eqnarray}
 R_{\alpha\mu\beta\nu} = C_{\alpha\mu\beta\nu} - \frac{R}{6} \left( g_{\alpha\beta} g_{\mu\nu} - g_{\alpha\nu} g_{\mu\beta} \right) + \frac{1}{2} \left( R_{\alpha\beta} g_{\mu\nu} + R_{\mu\nu} g_{\alpha\beta} - R_{\alpha\nu} g_{\beta\mu} - R_{\beta\mu} g_{\alpha\nu} \right)\,,
\end{eqnarray} 
where $C_{\alpha\mu\beta\nu}$ is the Weyl tensor.
\\\\
Let us start with the simplest case when the Weyl tensor is zero and the Ricci tensor is given by the cosmological constant
\begin{eqnarray}
C_{\alpha\mu\beta\nu} = 0\,,\, R_{\mu\nu}= \Lambda g_{\mu\nu}\,.
\end{eqnarray} 
Then the Riemann tensor can be decomposed into the components
\begin{eqnarray}
 R_{\alpha\mu\beta\nu} = \frac{\Lambda}{3} \left( g_{\alpha\beta} g_{\mu\nu} - g_{\alpha\nu} g_{\mu\beta} \right)\,,
 \label{Rmaxsym}
\end{eqnarray} 
which means that we consider de Sitter space-time ($\Lambda>0$) or anti-de Sitter space-time ($\Lambda<0$) \cite{Nastase:2015wjb}. Our further analysis takes place in a state of global equilibrium (\ref{killingeq}), so we choose static patch of dS space-time, therefore having a time-like Killing vector.
\\\\
Now we construct the hydrodynamic gradient expansion for the current. The only possibility to add space-time curvature to the current expression is the scalar curvature $R$, which is of the second order in gradients. Then to make this term pseudovector one has to multiply it by the axial quantity $w_{\mu}$, which also contains a derivative and therefore the axial current will be of the third order. Thus, the curvature effects arise only in the third order in gradients. Since the Ricci tensor is proportional to the cosmological constant, this term will have the form $\Lambda w_{\mu}$
\begin{eqnarray}
 j_{\mu}^{A(3)} = \xi_{1}(T)w^{2}w_{\mu} + \xi_{2}(T) \alpha^{2}w_{\mu} + \xi_{3}(T)(\alpha w)\alpha_{\mu} + \xi_{\Lambda}(T) \Lambda w_{\mu}\,.
 \label{current}
\end{eqnarray} 
We will denote the coefficient before the term with the cosmological constant as $\xi_{\Lambda}(T)$. The unknown coefficients $\xi_{n} (T)$ and $\xi_{\Lambda}(T)$ depend on the proper
temperature $T$. 

\subsection{Axial current divergence}
Due to the symmetry of the chosen Riemann tensor (the Weyl tensor is zero), there is no gravitational anomaly, and the axial current in this case is conserved
\begin{eqnarray}
    \nabla_{\mu} j^{\mu}_{A} = 0\,.
    \label{divcurr}
\end{eqnarray} 
The covariant derivatives of the kinematic variables can be found  from (\ref{killingeq}), (\ref{doubleder}) and the condition $ u^{\mu} u_{\mu} = 1 $
\begin{eqnarray}
\begin{cases}
\nabla_{\mu} T = T^{2} \alpha_{\mu}\,, \\
\nabla_{\mu}w_{\nu} = T \left( -g_{\mu\nu} (\alpha w) + \alpha_{\mu}w_{\nu} \right)\,, \\
\nabla_{\mu}\alpha_{\nu} = T \left( w^{2} \left( g_{\mu\nu} - u_{\mu} u_{\nu} \right) - \alpha^{2} u_{\mu} u_{\nu} -w_{\mu} w_{\nu} - u_{\mu} \eta_{\nu} - u_{\nu} \eta_{\mu} \right) - \frac{1}{T} \frac{\Lambda}{3} \left( g_{\mu\nu} - u_{\mu} u_{\nu} \right)\, ,  
\end{cases}
\label{covderivatives}
\end{eqnarray} \newline
where $\eta_{\mu} = \epsilon_{\mu\nu\rho\sigma} u^{\nu} w^{\rho} \alpha^{\sigma} $.
\\\\
An important consequence of considering the system in global thermodynamic equilibrium is that hydrodynamics is defined only in terms of the first gradients of velocity (acceleration and vorticity), and all higher derivatives are expressed through the lower ones. This is clearly seen from the system (\ref{covderivatives}). Now, substituting (\ref{current}) into (\ref{divcurr}), we obtain an expression for the axial current divergence 
\begin{eqnarray}
\label{axialdiv}
 \nabla_{\mu} j^{\mu (3)}_{A} &=& (\alpha w)w^{2} \left( T^{2}\xi_{1}' - 3T\xi_{1} + 2T\xi_{3} \right) +(\alpha w)\alpha^{2} \left( T^{2}\xi_{2}' - 3T\xi_{2} + T^{2}\xi_{3}' - T\xi_{3} \right) \nonumber \\
&&+ \Lambda (\alpha w) \left( T^{2} \xi_{\Lambda}' - 3T \xi_{\Lambda} - \frac{4}{3T} \xi_{3} - \frac{2}{3T} \xi_{2} \right)  = 0\,.
\end{eqnarray} 
The pseudo-scalar terms before each bracket in this expression are independent, so we can equate all expressions in brackets to zero and obtain a system of differential equations
\begin{eqnarray}
 \begin{cases}
   T^{2}\xi_{1}' - 3T\xi_{1} + 2T\xi_{3}  = 0\,, \\
   T^{2}\xi_{2}' - 3T\xi_{2} + T^{2}\xi_{3}' - T\xi_{3} = 0\,, \\
   T^{2} \xi_{\Lambda}' - 3T \xi_{\Lambda} - \frac{4}{3T} \xi_{3} - \frac{2}{3T} \xi_{2} = 0\,.
 \end{cases}
 \end{eqnarray} 
Based on dimensional analysis (in a massless theory without boundary conditions, etc., temperature $T$ is the only dimensional parameter), we move on to the  dimensionless coefficients $\lambda$
\begin{eqnarray}
\label{redefine1}
 \xi_{1} = T^{3} \lambda_{1}, \quad \xi_{2} = T^{3} \lambda_{2}, \quad \xi_{3} = T^{3} \lambda_{3}, \quad \xi_{\Lambda} = T \lambda_{\Lambda}\,.  
\end{eqnarray} 
And left with the system ($\lambda_{1}$ remains arbitrary)
\begin{equation}
 \begin{cases} 
   \lambda_{3}  = 0\,, \\
   \lambda_{\Lambda} = - \frac{\lambda_{2}}{3}\,.
 \end{cases}
 \label{system}
\end{equation} 
The first equality in (\ref{system}) provides the conservation of the axial current in flat space-time in the absence of external fields \cite{Prokhorov:2018bql}.
\\\\
The phenomena analyzed in this paper, from the point of view of conservation laws, can also be derived directly from the equilibrium perturbation theory \cite{Buzzegoli:2017cqy, Prokhorov:2018bql}, where they are determined by the effective interaction with boost operator (in the case of acceleration) and with operator of angular momentum (in the case of vorticity). In particular, in this way the coefficients $\lambda_1$, $\lambda_2$ and $\lambda_3$ were obtained in flat space-time for massless spin 1/2 field (\ref{diraccurrent12}).
The key is the second of the relations (\ref{system}). The factor $\lambda_{\Lambda}$ cannot be obtained in the same way from the density operator in flat space-time, but we can obtain it simply using the duality (\ref{system}) from the flat space transport coefficient $\lambda_2$. In particular, using (\ref{diraccurrent12}) and (\ref{system}), we obtain the following expression for the current of massless Dirac fields
\begin{eqnarray}
    \label{fermionacurrent1}
    j_{A}^{\mu(3)} = \left( \ -\frac{\omega^{2}}{24\pi^{2}} - \frac{a^{2}}{8\pi^{2}} + \frac{R}{96\pi^{2}} \right)\omega^{\mu}\,,
\end{eqnarray} 
where we used $ \Lambda = \frac{R}{4} $.
Thus, we obtained a new term in the axial current proportional to the scalar curvature in the hydrodynamic gradient expansion.  This current has the same conductivity as the acceleration term, according to (\ref{system}), which adds an additional element to the hydrodynamic/gravitational duality. We will also return to this issue in the Discussion section. The same expression for the axial current in AdS space-time was obtained by another method in \cite{Ambrus:2021eod}.

\section{Anomalous transport: gravitational anomaly and cosmological constant}
\label{sec 3}
\subsection{Riemann tensor and gradient expansion}
In this section we consider the more general case of space-time with a non-zero Weyl tensor and generalize the derivation of \cite{Prokhorov:2022udo} to the case with a non-zero Ricci tensor proportional to the cosmological constant. Our starting point will be a general decomposition for the Riemann tensor \cite{Prokhorov:2022udo}
\begin{eqnarray}
   R_{\mu\nu\alpha\beta} &=& u_{\mu} u_{\alpha} A_{\nu\beta} + u_{\nu} u_{\beta} A_{\mu\alpha} - u_{\nu} u_{\alpha} A_{\mu\beta} - u_{\mu} u_{\beta} A_{\nu\alpha} + \epsilon_{\mu\nu\lambda\rho}u^{\rho} \left( u_{\alpha} B^{\lambda}_{\:\beta} - u_{\beta}B^{\lambda}_{\:\alpha} \right) + \nonumber \\
&& + \epsilon_{\alpha\beta\lambda\rho}u^{\rho} \left( u_{\mu} B^{\lambda}_{\:\nu} - u_{\nu}B^{\lambda}_{\:\mu} \right) + \epsilon_{\mu\nu\lambda\rho}\epsilon_{\alpha\beta\eta\sigma} u^{\rho} u^{\sigma} C^{\lambda\eta}\,.
\label{riemann}
\end{eqnarray} 
The tensors $A_{\mu\nu}$, $B_{\mu\nu}$ and $C_{\mu\nu}$ are covariant generalizations of similar three-dimensional tensors from \cite{Landau:1975pou}
\begin{equation}
    \begin{cases}
        A_{\mu\nu} = u^{\alpha} u^{\beta} R_{\alpha\mu\beta\nu}\,, \vspace{0.2 cm} \\
        B_{\mu\nu} = \frac{1}{2} \epsilon_{\alpha\mu\eta\rho} u^{\alpha} u^{\beta} R_{\quad\beta\nu}^{\eta\rho}\,, \vspace{0.2 cm} \\ 
        C_{\mu\nu} = \frac{1}{4} \epsilon_{\alpha\mu\eta\rho} \epsilon_{\beta\nu\lambda\gamma} u^{\alpha} u^{\beta} R^{\eta\rho\lambda\gamma}\,.
    \end{cases}
    \label{ABC1}
\end{equation}
These tensors have properties 
\begin{eqnarray}
  A_{\mu\nu} = A_{\nu\mu},  \quad  C_{\mu\nu} = C_{\nu\mu}\,,  \quad B^{\mu}_{\mu}=0\,, \quad
A_{\mu\nu}u^{\nu} = C_{\mu\nu}u^{\nu} = B_{\mu\nu}u^{\nu} = B_{\nu\mu}u^{\nu} = 0\,.
\label{properties}
\end{eqnarray} 
The tensors $A_{\mu\nu}$ and $C_{\mu\nu}$ can be decomposed into a traceless and a trace parts, and the pseudotensor $B_{\mu\nu}$ can be decomposed into symmetric and anti-symmetric parts
\begin{eqnarray}
  A_{\mu\nu} &=& \widetilde{A}_{\mu\nu} + \frac{A}{3} \left( g_{\mu\nu} -u_{\mu}u_{\nu} \right)\,, \nonumber \\
  C_{\mu\nu} &=& \widetilde{C}_{\mu\nu} + \frac{C}{3} \left( g_{\mu\nu} -u_{\mu}u_{\nu} \right)\,, \nonumber \\
  B_{\mu\nu} &=& B^{s}_{\mu\nu} +B^{a}_{\mu\nu}\,,
 \label{newcond}
\end{eqnarray} 
accordingly, we have
\begin{eqnarray}
&&  \widetilde{A}_{\mu}^{\mu} = \widetilde{C}_{\mu}^{\mu} = 0\,,  \quad  \widetilde{A}_{\mu\nu}u^{\nu} = \widetilde{C}_{\mu\nu}u^{\nu} =B^{s}_{\mu\nu}u^{\nu}=B^{a}_{\mu\nu}u^{\nu}= 0\,, \nonumber \\
&&  A_{\mu}^{\mu} = A\,,  \quad  C_{\mu}^{\mu} = C\, , \quad B^{s}_{\mu\nu}=B^{s}_{\nu\mu}\,,\quad 
  B^{a}_{\mu\nu}=-B^{a}_{\nu\mu}\,.
 \label{newprop}
\end{eqnarray} 
Ricci tensor has the form
\begin{eqnarray}
   R_{\mu\nu} = g^{\alpha\beta} R_{\alpha\mu\beta\nu} = u_{\mu}u_{\nu}\left( A+C \right) + \left( A_{\mu\nu}+C_{\mu\nu} \right) - Cg_{\mu\nu} + B^{\lambda\alpha} u^{\rho} \left( u_{\nu} \epsilon_{\alpha\lambda\mu\rho} + u_{\mu}\epsilon_{\alpha\lambda\nu\rho} \right)\,. 
   \label{ricci}
\end{eqnarray} 
To illustrate the new elements, we now consider two cases: the Ricci tensor is equal to zero (the case considered in \cite{Prokhorov:2022udo}), and the Ricci tensor is equal to the cosmological constant (present case).
\\\\
\textit{1) Ricci tensor is zero, Weyl tensor is not zero: $R_{\mu\nu}=0, \, C_{\mu\nu\alpha\beta}\neq 0$.}\\

The condition $R_{\mu\nu}=0$ leads to additional properties for the tensors $A_{\mu\nu},B_{\mu\nu}$ and $C_{\mu\nu}$ such as (compare with \cite{Landau:1975pou})
\begin{eqnarray}
    A_{\mu\nu} = -C_{\mu\nu}\,, \quad A^{\mu}_{\mu} = 0\,, \quad B_{\mu\nu} = B_{\nu\mu}\,.
    \label{conditions}
\end{eqnarray}
This means that
\begin{eqnarray}
A=C=0\,,\quad B^a_{\mu\nu}=0\,,\quad \widetilde{A}_{\mu\nu} = -\widetilde{C}_{\mu\nu}\,.
    \label{conditions1}
\end{eqnarray}
Then, writing the axial current in terms of all possible pseudo-vectors arising in the third order of gradients, we obtain
\begin{eqnarray}
   j_{\mu}^{A(3)} = \xi_{1}(T)w^{2}w_{\mu} + \xi_{2}(T) \alpha^{2}w_{\mu} + \xi_{3}(T)(\alpha w)\alpha_{\mu} + \xi_{4}(T) \widetilde{A}_{\mu\nu}w^{\nu}  + \xi_{5}(T)B^{s}_{\mu\nu}\alpha^{\nu}\,.
   \label{dec1}
\end{eqnarray}
By considering covariant derivative of the (\ref{dec1}) and considering that it should lead to anomaly (\ref{gravanomaly}), a key relation (\ref{key}), fixing duality with gravitational anomaly was obtained \cite{Prokhorov:2022udo}. 
\\\\
We note that there could be some extra terms in the third order, including, for example
\begin{eqnarray}
\nabla_{\lambda} \widetilde{A}_{\mu\nu}\,, \quad
  \nabla_{\lambda} B^s_{\mu\nu}\,, \quad
\nabla_{\lambda} R_{\mu\nu\rho\sigma} \epsilon^{\alpha_{1}\alpha_{2}\alpha_{3}\alpha_{4}}\,,  \quad 
 \nabla_{\lambda} R_{\mu\nu\rho\sigma} \epsilon^{\alpha_{1}\alpha_{2}\alpha_{3}\alpha_{4}} u^{\alpha} u^{\beta}\,, \quad
 \alpha_{\lambda} R_{\mu\nu\rho\sigma} \epsilon^{\alpha_{1}\alpha_{2}\alpha_{3}\alpha_{4}}\,.
 \label{zeroderivatives}
\end{eqnarray} 
But all the index contractions which makes these terms pseudovectors will be equal zero because of Bianchi identities
\begin{eqnarray}
     R_{\mu\nu\rho\sigma} + R_{\mu\sigma\nu\rho} + R_{\mu\rho\sigma\nu} &=& 0\,, \nonumber \\   
     \nabla_{\lambda} R_{\mu\nu\rho\sigma} +  \nabla_{\sigma} R_{\mu\nu\lambda\rho} + \nabla_{\rho} R_{\mu\nu\sigma\lambda}  &=& 0\,,
\end{eqnarray}
and the condition $\nabla_{\lambda} R_{\mu\nu} = 0$.  In the case $R_{\mu\nu} = 0$, considered in \cite{Prokhorov:2022udo}, this condition is satisfied. However, it is also valid in the more general case $R_{\mu\nu} = \Lambda g_{\mu\nu}$ due to the metricity condition $\nabla_{\lambda} g_{\mu\nu} = 0$. Also there is possibility to add another type of terms like 
\begin{eqnarray}
  \left( \nabla_{\mu} \alpha_{\nu} \right) \omega^{\nu}\,,  \quad  \left( \nabla_{\mu} \omega_{\nu} \right) \alpha^{\nu}\,,  \quad
 \left( \nabla \alpha \right) \omega^{\mu}\,,  \quad \left( \nabla \omega \right) \alpha^{\mu}\,,
 \label{pointlessderivatives}
\end{eqnarray} 
but they can be transformed into terms already contained in the expression (\ref{dec1}) for the axial current.
\\\\
\textit{2) Ricci tensor is given by the cosmological constant, Weyl tensor is not zero: $R_{\mu\nu}=\Lambda g_{\mu\nu}, C_{\mu\nu\alpha\beta}\neq 0$.}\\

Now fixing $R_{\mu\nu} = \Lambda g_{\mu\nu}$ doesn't affect the condition  $\nabla_{\lambda} R_{\mu\nu} = 0$, and the conditions mentioned above don't change either and there will be no extra terms such as (\ref{zeroderivatives}) and (\ref{pointlessderivatives}) in the current. But the equations (\ref{conditions}) and (\ref{conditions1}) are no longer valid and the tensors have the form (\ref{newcond}). Therefore, using (\ref{ricci}), (\ref{newprop}) and $ R_{\mu\nu} = \Lambda g_{\mu\nu} $ we can write the following equations and define the constrains and symmetries of tensors $A_{\mu\nu}$, $B_{\mu\nu}$ and $C_{\mu\nu}$
\begin{equation}
 \begin{cases}
    u^{\mu}u^{\nu}R_{\mu\nu} = \Lambda = \frac{R}{4} = A \,,  \\
    g^{\mu\nu} R_{\mu\nu} = 4 \Lambda = 4A = 2(A-C) \Rightarrow A = -C\,,  \\
    u^{\mu}R_{\mu\nu} = \Lambda u_{\nu} = -Cu_{\nu} + B^{\lambda\alpha}\varepsilon_{\alpha\lambda\nu\rho}u^{\rho} \Rightarrow B^{\lambda\alpha} = B^{\alpha\lambda}\,.
 \end{cases}
\end{equation}
Substituting these expressions into the Ricci tensor (\ref{ricci}), we obtain the final conditions for the tensors $A_{\mu\nu}$, $B_{\mu\nu}$ and $C_{\mu\nu}$
\begin{equation}
 \begin{cases}
    A_{\mu\nu} = - C_{\mu\nu}\,, \\
    B_{\alpha\lambda} = B_{\lambda\alpha} = B^s_{\lambda\alpha} \quad (B^a_{\lambda\alpha} = 0)\,,\\
    \Lambda = \frac{R}{4} = A = -C\,.
 \end{cases}
\end{equation}
As can be seen, the tensors $ A_{\mu\nu} $ and $ C_{\mu\nu} $ are related, so we don't need to add a term with tensor $ C_{\mu\nu} $ to the expression for the current. Since we have separated the traceless part from the tensor $ A_{\mu\nu} $, a new term proportional to the cosmological constant  should appear in the current, since  $ A_{\mu\nu} \omega^{\nu} = \widetilde{A}_{\mu\nu} \omega^{\nu} + \Lambda \omega_{\mu} $. Therefore we have
\begin{eqnarray}
 j_{\mu}^{A(3)} = \xi_{1}(T)w^{2}w_{\mu} + \xi_{2}(T) \alpha^{2}w_{\mu} + \xi_{3}(T)(\alpha w)\alpha_{\mu} + \xi_{4}(T)\widetilde{A}_{\mu\nu}w^{\nu}  + \xi_{5}(T)B^s_{\mu\nu}\alpha^{\nu}  + \xi_{\Lambda}(T) \Lambda w_{\mu}\,.
 \label{finalcurrent}
\end{eqnarray}

\subsection{Axial current divergence}

Due to the quantum anomaly, the axial current isn't conserved in the presence of external gravitational fields. The gravitational chiral anomaly (\ref{gravanomaly}) can be expressed through the tensors $ \widetilde{A}_{\mu\nu} $ and $ B^s_{\mu\nu} $
\begin{eqnarray}
\label{divergence}
\nabla_{\mu} j_{A}^{\mu}  =\mathcal{N} \epsilon^{\mu\nu\alpha\beta} R_{\mu\nu\lambda\rho} R_{\alpha\beta}^{\quad\lambda\rho} = 16 \mathcal{N}  \left( A^{\mu\nu} - C^{\mu\nu} \right) B^s_{\mu\nu} = 32 \mathcal{N}  \widetilde{A}^{\mu\nu} B^s_{\mu\nu}\,. 
\end{eqnarray}
In the case under consideration with a nonzero Weyl tensor, instead of (\ref{covderivatives}), we obtain
\begin{equation}
 \begin{cases}
    \nabla_{\mu} T = T^{2} \alpha_{\mu}\,, \\
    
    \nabla_{\mu} u_{\nu} = T\left( \epsilon_{\mu\nu\alpha\beta} u^{\alpha} w^{\beta} + u_{\mu}\alpha_{\nu} \right)\,, \\
    
    \nabla_{\mu}w_{\nu} = T \left( -g_{\mu\nu} (\alpha w ) + \alpha_{\mu}w_{\nu} \right) + \frac{1}{T} B^s_{\mu\nu}\,, \\

    \nabla_{\mu}\alpha_{\nu} = T \left( w^{2} \left( g_{\mu\nu} - u_{\mu} u_{\nu} \right) - \alpha^{2} u_{\mu} u_{\nu} -w_{\mu} w_{\nu} - u_{\mu} \eta_{\nu} - u_{\nu} \eta_{\mu}     \right) - \frac{1}{T} \left( \widetilde{A}_{\mu\nu} + \frac{\Lambda}{3} \left( g_{\mu\nu} - u_{\mu} u_{\nu} \right) \right)\,, \\

    \nabla^{\mu} \left( \widetilde{A}_{\mu\nu} w^{\nu} \right) = -3TB^s_{\mu\nu}w^{\mu} w^{\nu} + \frac{1}{T} \widetilde{A}^{\mu\nu} B^s_{\mu\nu}\,, \\

    \nabla^{\mu} \left( B^s_{\mu\nu} \alpha^{\nu} \right) = 3T \widetilde{A}_{\mu\nu} w^{\mu} \alpha^{\nu} - \frac{1}{T} \widetilde{A}^{\mu\nu} B^s_{\mu\nu} -  T B^s_{\mu\nu} w^{\mu} w^{\nu} - T B^s_{\mu\nu} \alpha^{\mu} \alpha^{\nu}\,.  
 \end{cases}
\end{equation} 
Now for clarity, let us write out the derivatives of each term of the axial current (\ref{finalcurrent})
\begin{equation}
    \begin{cases}
\nabla_{\mu} \left( \xi_{1} w^{2}w^{\mu} \right) = (\alpha w)w^{2} \left( T^{2} \xi_{1}' - 3T\xi_{1} \right) + B^s_{\mu\nu} w^{\mu}w^{\nu} \left(  \frac{2}{T} \xi_{1} \right)\,,
\\ 
\nabla_{\mu} \left( \xi_{2} \alpha^{2}w^{\mu} \right) = (\alpha w)\alpha^{2} \left( T^{2} \xi_{2}' - 3T\xi_{2} \right) + \widetilde{A}_{\mu\nu} \alpha^{\mu} w^{\nu} \left(- \frac{2}{T} \xi_{2} \right) +  \Lambda (\alpha w) \left(- \frac{2}{3T} \xi_{2} \right)\,,
\\
\nabla_{\mu} \left( \xi_{3} (\alpha w)\alpha^{\mu} \right) = (\alpha w) 
\Bigg[ \alpha^{2} \left( T^{2} \xi_{3}' - T\xi_{3} \right) + w^{2} \left( 2T \xi_{3} \right) + \Lambda \left( -\frac{4}{3T} \xi_{3} \right) \Bigg] + \widetilde{A}_{\mu\nu} \alpha^{\mu} w^{\nu} \left( -\frac{1}{T} \xi_{3} \right) + B^s_{\mu\nu} \alpha^{\mu} \alpha^{\nu} \left(  \frac{1}{T} \xi_{3} \right)\,,
\\
\nabla^{\mu} \left( \xi_{4}\widetilde{A}_{\mu\nu}w^{\nu} \right) =  \widetilde{A}_{\mu\nu} \alpha^{\mu} w^{\nu} \left( \xi_{4}' T^{2} \right) + B^s_{\mu\nu} w^{\mu} w^{\nu} \left( - 3T \xi_{4} \right) + \widetilde{A}^{\mu\nu} B^s_{\mu\nu} \left(  \frac{1}{T} \xi_{4} \right)\,,
\\
\nabla^{\mu} \left( \xi_{5}B^s_{\mu\nu}\alpha^{\nu} \right) = B^s_{\mu\nu} \alpha^{\mu} \alpha^{\nu} \left( T^{2} \xi_{5}' - T \xi_{5} \right) + B^s_{\mu\nu} w^{\mu} w^{\nu} \left( - T \xi_{5} \right) + \widetilde{A}^{\mu\nu} B^s_{\mu\nu} \left( - \frac{1}{T} \xi_{5} \right) + \widetilde{A}_{\mu\nu} \alpha^{\mu} w^{\nu} \left( 3T \xi_{5} \right)\,, 
\\
\nabla^{\mu} \left( \xi_{\Lambda} \Lambda w_{\mu} \right) = \Lambda (\alpha w) \left( T^{2} \xi_{\Lambda}' - 3T \xi_{\Lambda} \right)\,.    
    \end{cases}
 \label{final derivatives}
\end{equation}
Combining (\ref{divergence}) and (\ref{final derivatives}) we obtain
\begin{eqnarray}
\nabla_{\mu} j^{\mu (3)}_{A} &=& (\alpha w)w^{2} \left( T^{2}\xi_{1}' - 3T\xi_{1} + 2T\xi_{3} \right) + (\alpha w)\alpha^{2} \left( T^{2}\xi_{2}' - 3T\xi_{2} + T^{2}\xi_{3}' - T\xi_{3} \right) + \widetilde{A}^{\mu\nu} B^s_{\mu\nu} \left(  \frac{1}{T} \xi_{4} - \frac{1}{T} \xi_{5} \right) +
\nonumber \\
&&+ \widetilde{A}_{\mu\nu}\alpha^{\mu}w^{\nu} \left( -\frac{2}{T} \xi_{2} - \frac{1}{T} \xi_{3} + 3T\xi_{5} + T^{2} \xi_{4}' \right) +
B^s_{\mu\nu}w^{\mu}w^{\nu} \left(  \frac{2}{T} \xi_{1} -3T \xi_{4} -T \xi_{5} \right) +
\nonumber \\
&& + B^s_{\mu\nu} \alpha^{\mu} \alpha^{\nu} \left( T^{2} \xi_{5}' -T \xi_{5} + \frac{1}{T} \xi_{3} \right) + \Lambda (\alpha w) \left( T^{2} \xi_{\Lambda}' - 3T \xi_{\Lambda} - \frac{4}{3T} \xi_{3} - \frac{2}{3T} \xi_{2} \right)\nonumber \\ 
&=& 32 \mathcal{N} \widetilde{A}^{\mu\nu} B^s_{\mu\nu}\,.
\end{eqnarray}
As in (\ref{axialdiv}) all pseudo-scalars are independent, so we can write a system of equations 
\begin{equation}
 \begin{cases}
 
    T^{2}\xi_{1}' - 3T\xi_{1} + 2T\xi_{3}  = 0\,, \\
    T^{2}\xi_{2}' - 3T\xi_{2} + T^{2}\xi_{3}' - T\xi_{3} = 0\,, \\
  - \frac{2}{T} \xi_{2} - \frac{1}{T} \xi_{3} + 3T\xi_{5} + T^{2} \xi_{4}' = 0\,, \\
     \frac{2}{T} \xi_{1} -3T \xi_{4} -T \xi_{5} = 0\,, \\
    T^{2} \xi_{5}' -T \xi_{5} + \frac{1}{T} \xi_{3} = 0\,, \\
     \frac{1}{T} \xi_{4} - \frac{1}{T} \xi_{5}  = 32 \mathcal{N}\,, \\
    T^{2} \xi_{\Lambda}' - 3T \xi_{\Lambda} - \frac{4}{3T} \xi_{3} - \frac{2}{3T} \xi_{2} = 0\,.

 \end{cases}
\end{equation}
By redefining dimensionless constants as we did it in (\ref{redefine1}), we obtain 
\begin{eqnarray}
    \label{redefine2}
    \xi_{1} = T^{3} \lambda_{1}\,, \quad \xi_{2} = T^{3} \lambda_{2}\,, \quad \xi_{3} = T^{3} \lambda_{3}\,, \quad
\xi_{4} = T \lambda_{4}\,, \quad \xi_{5} = T \lambda_{5}\,, \quad \xi_{\Lambda} = T \lambda_{\Lambda}\,.
\end{eqnarray}
Then
\begin{equation}
 \begin{cases}
 
    \lambda_{3}  = 0\,, \\
   -2\lambda_{2} + 3\lambda_{5} + \lambda_{4} = 0\,, \\
   2\lambda_{1} - 3\lambda_{4} -\lambda_{5} = 0\,, \\
    \lambda_{4} - \lambda_{5} = 32 \mathcal{N}\,,  \\
    -2\lambda_{\Lambda} - \frac{2}{3} \lambda_{2} = 0\,. \\
   
 \end{cases}
\end{equation}
From here we can finally write the solution
\begin{eqnarray}
    \label{relationships}
   \lambda_{1} - \lambda_{2} = 32 \mathcal{N}, \quad \lambda_4 = \frac{\lambda_1}{2} + 8 \mathcal{N}, \quad \lambda_5 = \frac{\lambda_1}{2} - 24 \mathcal{N}, \quad \lambda_{\Lambda} = -\frac{\lambda_{2}}{3}\,.
\end{eqnarray}
The first three equalities in (\ref{relationships}) are consistent with the work \cite{Prokhorov:2022udo}, while the last one tells us that the coefficients between the cosmological constant term the squared acceleration term square are related as in the simpler case with the zero Weyl tensor (\ref{system}). Thus, if we know the gravitational chiral anomaly, then the current in third order in gradients is determined to within one arbitrary coefficient. It is convenient to choose $\lambda_1$ as this arbitrary coefficient, since it can be calculated in flat space-time. As a result, knowing only one transport coefficient in flat space-time, we complete the full expression for the current in curved space in the third order. As a result, we have for the general case
\begin{eqnarray} \nonumber
j^{A (3)}_{\mu} &=&  \left(\lambda_1 \omega^{2} +(32 \mathcal{N}-\lambda_1)\left[\frac{R}{12}-a^{2}\right]\right) \omega_{\mu}+
    \left(8\mathcal{N}+\frac{\lambda_1}{2}\right)R_{\alpha\mu\beta\nu}u^{\alpha}u^{\beta}\omega^{\nu}+ \\
    &&+
    \left(\frac{\lambda_1}{4}-12 \mathcal{N} \right)R^{\eta\rho}_{\quad\beta\nu}\epsilon_{\alpha\mu\eta\rho} u^{\alpha} u^{\beta}  a^{\nu}\,,
     \label{full1}
\end{eqnarray}
where we moved from the tensors $A_{\mu\nu}$ and $B_{\mu\nu}$  to the Riemann tensor by means of (\ref{ABC1}). 
\\\\
Formula (\ref{full1}) is our main result. It describes the axial current of massless particles in a vortical and accelerated medium in global thermodynamic equilibrium in curved space with a nonzero Weyl tensor, and the Ricci tensor of the form (\ref{rl}). One can also assume that by adding also the linear order (\ref{cve 12}), the formula will be accurate (all higher orders will be zero) at sufficiently high temperatures\footnote{This polynomiality is a fairly rare and interesting phenomenon for quantum field theory. The absence of higher order terms has been directly shown in a number of approaches for specific cases \cite{Prokhorov:2019hif, Palermo:2021hlf, Becattini:2020qol, Stone:2018zel} and is characteristic only for the massless theory. Formally, the absence of higher order terms corresponds to the absence of negative powers of temperature. At low temperatures, however, the phase transitions associated with the singularity of modes at the horizon should be expected \cite{Prokhorov:2023dfg} at least for half integer spins, which will modify the obtained formulas.}.
\\\\
In the particular case of massless Dirac fields, we know the anomaly $\mathcal{N}=1/(384 \pi^2)$ (\ref{gravanomaly12a}) and the coefficient $\lambda_1 = -1/(24 \pi^2)$  (\ref{diraccurrent12}). As a result, the current will have the form
\begin{eqnarray} \label{full}
    j_{\mu}^{A(3)} = \left( - \frac{\omega^{2}}{24\pi^{2}} + \frac{1}{8\pi^{2}}\left[\frac{R}{12} -a^{2}\right]\right)\omega_{\mu} -
    \frac{1}{24\pi^2} R^{\eta\rho}_{\quad\beta\nu}\epsilon_{\alpha\mu\eta\rho} u^{\alpha} u^{\beta}  a^{\nu}\,,
\end{eqnarray}
which generalizes the formulas (\ref{diraccurrent12}) and (\ref{fermionacurrent1}).
\section{Discussion}
\label{sec 4}

\subsection{Thermodynamic equilibrium}

From the very beginning we used the assumption that the system is in the so-called global thermodynamic equilibrium, which corresponds to the entropy maximum (as well as invariance of the density operator with respect to the choice of hypersurface) \cite{Becattini:2016stj}. Beyond this equilibrium our results may be unfair (or give only zero order contribution). We will discuss the global thermodynamic equilibrium in more detail.
\\\\
A consequence of this equilibrium is that the inverse temperature vector should be the Killing vector  (\ref{killingeq}). Since its square $\beta_{\mu}\beta^{\mu}=T^{-2}>0$ is related to the temperature, this vector should be time-like and also future-directed, since $\beta_0=u_{0}/T>0$ \cite{Becattini:2017ljh}.
The spaces containing timelike Killing vector are well known stationary spaces. In particular, any static space-time whose components do not depend on time is stationary. Thus, our derivation, is valid only for stationary spaces. In  Section \ref{sec 2} we considered the case of (A)dS spaces. For (A)dS space we can choose a static metric
\begin{eqnarray}
    ds^2 = \left(1-\frac{\Lambda}{3}r^2\right)dt^2 - \frac{1}{1-\frac{\Lambda}{3}r^2} dr^2 - r^2\left(d\Theta^2+\sin^2\Theta \, d\phi^2\right)\,.
\end{eqnarray}
where the positive sign $\Lambda>0$ corresponds to the metrics of a static patch in de Sitter space-time, and $\Lambda<0$ corresponds to static coordinates of the anti-de Sitter space-time.
\\\\
Not all types of space-times are stationary and, accordingly, global equilibrium is not always possible. In particular, it is not obvious whether global equilibrium is possible in the expanding Universe, since, for example the Friedman metrics clearly depends on time (but local thermodynamic equilibrium should be possible \cite{Becattini:2022bia}). 
\\\\
The conditions of global equilibrium (\ref{killingeq}) are consistent with other known criteria of thermodynamic equilibrium. In particular, there is a well-known Tolman-Ehrenfest thermodynamic equilibrium criterion \cite{Tolman:1930zza, Bermond:2022mjo}, according to which the temperature in space with the metric $g_{\mu\nu}(\mathbf{x})$ is distributed according to the law
\begin{eqnarray}
T(\mathbf{x})=\frac{T_0}{\sqrt{g_{00}(\mathbf{x})}}\,,
\label{Tolman1}
\end{eqnarray}
which for stationary space-times can be written in covariant form \cite{Rovelli:2010mv, Bermond:2022mjo}
\begin{eqnarray}
T(\mathbf{x})=\frac{T_0}{\sqrt{g_{\mu\nu}(\mathbf{x})\xi^{\mu}(\mathbf{\mathbf{x}})\xi^{\nu}(\mathbf{x})}}\,,
\label{Tolman2}
\end{eqnarray}
where $\xi^{\mu}$ is a dimensionless Killing vector. It is clear, that (\ref{Tolman2}) is in full agreement with the global equilibrium, according to which the temperature is also defined in exactly the same way by the Killing vector as $\beta_{\mu}\beta^{\mu}\equiv g_{\mu\nu}\beta^{\mu}\beta^{\nu}=T^{-2}$. 
\\\\
Also, it is easy to see that in global equilibrium (\ref{killingeq}) the dissipative contribution to the stress-energy tensor vanishes, which also corresponds to equilibrium \cite{Luttinger:1964zz} 
\begin{eqnarray} \nonumber
T_{\mu\nu}^{\text diss} &=& -\eta (\nabla_{\mu} u_{\nu}+ \nabla_{\nu} u_{\mu} -u_{\mu} u^{\alpha}\nabla_{\alpha}u_{\nu} -u_{\nu} u^{\alpha}\nabla_{\alpha}u_{\mu} ) -\left( \zeta -\frac{2}{3}\eta\right)\nabla^{\alpha}u_{\alpha}(g_{\mu\nu}-u_{\mu}u_{\nu}) \\
&=& -\eta \left[ T (\nabla_{\mu}\beta_{\nu}+\nabla_{\nu}\beta_{\mu})+u_{\mu}\left(\frac{\nabla_{\nu} T}{T}-a_{\nu}\right)
+u_{\nu}\left(\frac{\nabla_{\mu} T}{T}-a_{\mu}\right)\right]+ \nonumber \\
&& +\left( \zeta -\frac{2}{3}\eta\right)\left(T\nabla_{\alpha}\beta^{\alpha}+\frac{\nabla_{\alpha} T}{T}u^{\alpha}\right)(g_{\mu\nu}-u_{\mu}u_{\nu})=0\,.
\label{diss}
\end{eqnarray}
The same, of course, applies to dissipative effects in a current.
\\\\
Thus, the conditions of global equilibrium look like a fairly general and natural assumption about the properties of the system. Note also that, generally speaking, equilibrium can lead to additional restrictions on the external field, as was in the case with the electromagnetic field \cite{Yang:2022ksq, Buzzegoli:2020ycf}. Such limitations are not significant in this paper, but we hope to explore them in the future.

\subsection{Qualitative description}

Current (\ref{full1}) (as well as (\ref{full})) does not explicitly depend on the properties of the medium, such as temperature or chemical potential, being poorly kinematic. This distinguishes it from CVE (\ref{cve 12}) or CME \cite{Zakharov:2012vv}. Therefore, they are associated with quantum vacuum fluctuations. Note that transport effects of vacuum kind are known (for example in external electromagnetic fields \cite{Chu:2018ksb} or there is well-known vacuum contribution to the stress-energy tensor in accelerated frame \cite{Sciama:1981hr}).
\\\\
This approach allows us to obtain an approximate qualitative picture for these effects. Consider for example the term with scalar curvature and acceleration in the current (\ref{full})
\begin{eqnarray}
     j_{\mu}^{A} = \frac{1}{2} \frac{|a|^{2}+R/12}{(2\pi)^2} \omega_{\mu}\,,
 \label{a2RwNew}
\end{eqnarray}
To understand at a qualitative level the meaning of (\ref{a2RwNew}), it is enough to take into account two phenomena. Firstly, according to the famous Unruh effect \cite{Unruh:1976db}, thermal radiation occurs in an accelerated medium with the Unruh temperature
\begin{eqnarray}
T_U = \frac{|a|}{2 \pi}\,.
 \label{TU}
\end{eqnarray}
Similarly, in de Sitter space the vacuum state is also characterized by a certain temperature \cite{Gibbons:1977mu}
\begin{eqnarray}
T_R = \frac{\sqrt{R/12}}{2 \pi}\,.
 \label{TR}
\end{eqnarray}
Temperatures (\ref{TU}) and (\ref{TR}) can be combined \cite{Deser:1997ri, Khakimov:2023emy} if we take into account, that the de Sitter space is embedded in a five-dimensional flat space. In this case, the radiation temperature is determined by the five-dimensional acceleration
\begin{eqnarray}
T_{aR}= \frac{|a_5|}{2 \pi} = \frac{\sqrt{|a|^2+R/12}}{2 \pi}\,,
 \label{TaR}
\end{eqnarray}
which is also true for anti-de Sitter space.
\\\\
On the other hand, we need to recall the well-known CVE (\ref{diraccurrent12}), according to which an axial current
\begin{eqnarray}
j^A_{\mu}= \frac{T^2}{6}  \omega_{\mu}\,,
 \label{cve1}
\end{eqnarray}
arises in a rotating (in flat space) medium at finite temperature.
CVE can be associated with the alignment of spins in a magnetic field, taking into account that rotation is similar to a magnetic field. Combining (\ref{cve1}) and (\ref{TaR}), that is, considering radiation in a vacuum as a medium with temperature (\ref{TaR}), we automatically obtain that a current should arise, which, up to the overall coefficient, coincides with (\ref{a2RwNew}). 
\\\\
Of course, the picture described is rough and is more intended to give an intuitive qualitative understanding of the effects occurring. And to be more precise, it is necessary to take into account all the peculiarities of the space with a horizon (which is not quite equivalent to a thermal medium \cite{Bunney:2023ipk}), which, in particular, is the reason for the difference in the overall coefficient.

\subsection{Duality between hydrodynamics and gravity}

The result we obtained continues to develop the entropy approach to gravity, which explains the effects of general relativity using the statistical properties \cite{Verlinde:2010hp}. Actually, the possibility of such a dual description suggests that the effects in gravitational fields should be accompanied by similar partner-phenomena within statistical approaches.
\\\\
There are a number of examples where such a relationship has been revealed, and the effects of gravity appear essentially in the limit of zero Newton's constant or zero curvature. This includes, for example, the well-known Unruh effect \cite{Unruh:1976db} - despite the fact that the accelerated system has zero curvature, it, like a black hole, is filled with thermal radiation bath. In this regard, we note a number of works \cite{Prokhorov:2019yft, Prokhorov:2019cik, Becattini:2017ljh}, where the Unruh effect was analyzed from the point of view of statistical approaches. Another example is the statistical derivation of the Einstein equation \cite{Jacobson:1995ab}, in which, for example, Newton's constant itself arises as the inverse entropy.
\\\\
Our result adds more elements to this duality. Indeed, on the one hand, there is an essentially gravitational effect - a gravitational chiral anomaly (\ref{gravanomaly}). Let us emphasize that this anomaly occurs only when space has non-zero curvature $R_{\mu\nu\alpha\beta}\neq 0$, while upon transition to a non-inertial system, the curvature remains equal to zero $R_{\mu\nu\alpha\beta}= 0$. However, we see, according to (\ref{relationships}), that this anomaly is ``hidden'' in the properties of the kinematic current (\ref{diraccurrent}) in an accelerated and vortical medium.
\\\\
We also note the relationship (\ref{system}) between the current $j^A_{\mu}\sim a^2 \omega_{\mu}$ in an accelerated vortical medium, and $j^A_{\mu}\sim R \omega_{\mu}$, in a vortical medium in the presence of a gravitational field, which indicates the similarity between scalar curvature $R$ and acceleration $a^2$.
\\\\
The statements made above can be made more concrete, if we remember both theories, within the framework of which the gravitational anomaly (\ref{gravanomaly}) and current (\ref{full}) can be obtained. The anomaly is associated with fundamental gravitational interaction. In the lowest order, by the definition, the corresponding vertex contains the stress-energy tensor
\begin{eqnarray} \label{dS}
    \delta S = -\frac{1}{2}\int d^4x \,\delta g_{\mu\nu} T^{\mu\nu}\,.
\end{eqnarray}
On the other hand, current (\ref{diraccurrent}) can be obtained using the statistical density operator \cite{Buzzegoli:2017cqy, Prokhorov:2018bql}
\begin{eqnarray}
\hat{\rho}=\frac{1}{Z}\exp\Big\{-\beta_{\mu}\hat{P}^{\mu}+\zeta  \hat{Q}
-\alpha_{\mu}\hat{K}^{\mu}_x-w_{\mu}\hat{J}^{\mu}_x
\Big\} \,,
\label{rho}
\end{eqnarray}
where $\zeta=\mu/T$, $\hat{P}^{\mu}$ is the four-momentum operator and $\hat{Q}$ is the charge operator. The first two terms correspond to the usual grand canonical distribution with , while the last two terms contain effective vertices, which describe statistically induced interaction, with the angular momentum operator $\hat{J}^{\mu}_x$, associated with vorticity $w_{\mu}=\omega_{\mu}/T$ and the boost operator $\hat{K}^{\mu}_x$, associated with acceleration $\alpha_{\mu}=a_{\mu}/T$ (the operators are shifted by the vector $x_{\mu}$). Moreover, (\ref{rho}) is exact for global thermodynamic equilibrium (\ref{killingeq}), which further illustrates the statistical nature of (\ref{rho}). Current (\ref{diraccurrent}) can be obtained within the framework of the equilibrium perturbation theory \cite{Buzzegoli:2017cqy}. For example, the term $\lambda_1 \omega^2\omega_{\mu}$ in (\ref{diraccurrent}) is given by a correlator of the form \cite{Prokhorov:2018bql, Prokhorov:2022snx}
\begin{eqnarray}
\lambda_1 =-\frac{1}{6}
\int_0^{|\beta|}d\tau_x d\tau_y d\tau_z
\langle T_{\tau} \hat{J}^{3}_{-i\tau_{x}}\hat{J}^{3}_{-i\tau_{y}}\hat{J}^{3}_{-i\tau_{z}}\hat{j}_A^{3}(0)\rangle_{\beta(x),c} \,,
\label{l1}
\end{eqnarray}
where the operators are shifted along the axis of the imaginary time along which integration is carried out from 0 to the inverse temperature $|\beta|=1/T$, and the index $\langle ...\rangle_{\beta(x),c} $ means that the averaging is taken using grand canonical distribution (i.e. using (\ref{rho}) with $\alpha_{\mu}=w_{\mu}=0$). Note that (\ref{rho}) and (\ref{l1}) are written in flat space-time. Moreover, they are written in ordinary Minkowski metrics. However, as we have shown in the main part of our work, the effects of the types (\ref{rho}) and (\ref{dS}) are related by the equations (\ref{relationships}).
\\\\
At the same time, our result can be interpreted as a confirmation of the equivalence principle, according to which, acceleration effects are equivalent to the effects in gravitational field. In most cases only the ``linear order'' is considered, that is, the equivalence at the level of acceleration and Christoffel symbols. Our derivation demonstrates the relationship between inertial effects and gravity at higher orders of metric and velocity derivatives.
\\\\
Thus, apparently, it can be argued that there is a class of theories that reproduce gravitational properties, despite the fact that they themselves do not contain gravity. Such theories can be considered as a limiting case when Newton's constant tends to zero. The role of such a limit is also evidenced by recent results on the algebra of observables \cite{Chandrasekaran:2022cip}.

\subsection{Quantum or classical?}

The conclusion we presented is completely unrelated to quantum theory calculations, which could create the illusion of the classical nature of the current (\ref{full1}). Evidently, this is not the case, since hydrodynamics, like thermodynamics, is a universal approach for describing both classical and quantum phenomena (just remember black hole radiation or superfluidity).
\\\\
The quantumness of (\ref{full}) becomes evident, if we analyse the way it can be directly obtained using the perturbation theory. In particular, $\omega^2\omega_{\mu}$ and $a^2\omega_{\mu}$ terms are expressed through loop diagrams of the form (\ref{l1}), which, of course, are elements of quantum theory.

\section{Conclusion}
In this paper we generalize the previously obtained kinematic vortical effect \cite{Prokhorov:2022udo} to the case of a nonzero Ricci tensor, proportional to the cosmological constant. We obtained not only the conservation of a previous result for the KVE, but also an entirely new term in the axial current, generated by the scalar curvature in a vortical fluid.  Analyzing the conservation of axial current, we relate this new term with the transport coefficient of the purely kinematic quadratic acceleration term. 
\\\\
We discover relationship between kinematic and gravitational effects, expressed in a system of equations connecting transport coefficients of two types, which indicates the gravitational-hydrodynamic duality and confirms the equivalence principle in the higher orders of gradient expansion. This relationship allows one by considering kinematic effects in a relativistic fluid (for example, in a quark-gluon plasma, in which sufficiently large vorticity and acceleration can be generated), to simulate quantum field effects in the gravitational field in the absence of real strong gravity.
\\\\
{\bf Acknowledgements}\\
The authors are thankful to Jian-Hua Gao for stimulating discussions. G.P. thanks Zuo-Tang Liang and Shandong University for the hospitality during the visit to Qingdao.
The work of G.P. and V.Z. was supported by Russian Science Foundation Grant No 24-22-00124. The work of V.Z. is partially supported by grant No. 0657-2020-0015 of the Ministry of Science and Higher Education of Russia.

\bibliography{lit}

\end{document}